\documentclass[twocolumn,prb,aps,superscriptaddress,showpacs]{revtex4}
\usepackage{amssymb}
\usepackage{dcolumn}
\usepackage{epsfig}
\usepackage{graphicx}
\usepackage{epstopdf}
\usepackage{float}
\usepackage[tbtags]{amsmath}

\newlength{\figwidth}
\newlength{\figwidthb}
\begin{document}

\title{Correlation of the Superconducting Critical Temperature with Spin and
Orbital Excitation Energies in (Ca$_{x}$La$_{1-x}$)(Ba$_{1.75-x}$La$%
_{0.25+x} $)Cu$_{3}$O$_{y}$ as Measured by Resonant Inelastic X-ray
Scattering}
\author{David Shai Ellis}
\affiliation{Physics Department, Technion-Israel Institute of Technology, Haifa 32000,
Israel}
\author{Yao-Bo Huang}
\affiliation{Swiss Light Source, Paul Scherrer Institut, CH-5232 Villigen PSI, Switzerland}
\affiliation{Beijing National Laboratory for Condensed Matter Physics, and Institute of
Physics, Chinese Academy of Sciences, Beijing 100190, China}
\author{Paul Olalde-Velasco}
\affiliation{Swiss Light Source, Paul Scherrer Institut, CH-5232 Villigen PSI, Switzerland}
\author{Marcus Dantz}
\affiliation{Swiss Light Source, Paul Scherrer Institut, CH-5232 Villigen PSI, Switzerland}
\author{Jonathan Pelliciari}
\affiliation{Swiss Light Source, Paul Scherrer Institut, CH-5232 Villigen PSI, Switzerland}
\author{Gil Drachuck}
\affiliation{Physics Department, Technion-Israel Institute of Technology, Haifa 32000,
Israel}
\author{Rinat Ofer}
\affiliation{Physics Department, Technion-Israel Institute of Technology, Haifa 32000,
Israel}
\author{Galina Bazalitsky}
\affiliation{Physics Department, Technion-Israel Institute of Technology, Haifa 32000,
Israel}
\author{Jorge Berger}
\affiliation{Department of Physics and Optical Engineering, ORT-Braude College, P.O. Box
78, 21982, Karmiel, Israel}
\author{Thorsten Schmitt}
\affiliation{Swiss Light Source, Paul Scherrer Institut, CH-5232 Villigen PSI, Switzerland}
\author{Amit Keren}
\email{keren@physics.technion.ac.il}
\affiliation{Physics Department, Technion-Israel Institute of Technology, Haifa 32000,
Israel}
\date{\today }

\begin{abstract}
Electronic spin and orbital ($dd$) excitation spectra of (Ca$_{x}$La$_{1-x}$%
)(Ba$_{1.75-x}$La$_{0.25+x}$)Cu$_{3}$O$_{y}$ samples are measured by
resonant inelastic x-ray scattering (RIXS). In this compound, $T_{c}$ of
samples with identical hole dopings is strongly affected by the Ca/Ba
substitution $x$ due to subtle variations in the lattice constants, while crystal symmetry and disorder as measured by line-widths are $x$ independent. We examine two extreme
values of $x$ and two extreme values of hole-doping content $y$ corresponding to  antiferromagnetic and superconducting states.  The $x$ dependence of the spin mode energies is approximately the same for both the antiferromagnetic
and superconducting samples.  This clearly demonstrates that RIXS is sensitive to $J$ even in doped samples.  A positive correlation
between the superexchange $J$ and the maximum of $T_{c}$ at optimal doping ($T_{c}^{max}$) is observed. 
We also measured the $x$ dependence of the $d_{xy}\rightarrow d_{x^{2}-y^{2}}$ and $d_{xz/yz}\rightarrow d_{x^{2}-y^{2}}$ orbital splittings. We infer that the effect of the unresolved $d_{3z^{2}-r^{2}}\rightarrow d_{x^{2}\rightarrow y^{2}}$ excitation on $T_{c}^{max}$ is much smaller than the effect of $J$.  There appears to be dispersion in the $d_{xy}\rightarrow d_{x^{2}-y^{2}}$ peak of up to 0.05 eV. Our fitting of the peaks furthermore indicates an asymmetric dispersion for the $d_{xz/yz}\rightarrow d_{x^{2}-y^{2}}$ excitation. A peak at $\sim $0.8 eV is also observed, and attributed to a $dd$ excitation in the chain layer.

\end{abstract}


\maketitle

\section{Introduction}

\label{sect:intro}

Theories built around coupling of the electron spins $\mathbf{S}$ have
become the prominent models for high-$T_{c}$ superconductivity\cite{Scalapino12}. A key parameter in these theories is the magentic superexchange
energy $J$, which is predicted to limit\cite{Anderson87} or set\cite{Scalapino98,Sushkov04} the critical temperature for superconductivity. One method of testing this has been to  compare $T_{c}$ against $J$ for a variety of cuprates\cite%
{Munoz00,Munoz02,Mallet13,Dean14}. The study of Munoz $et~al.$\cite{Munoz00}
resulted in a $\Delta T_{c}^{max}$/$\Delta J\sim $3 K/meV. However, if
the compounds vary in structures and nuances, other factors besides $J
$ are likely to influence the $J$-$T_{c}$ plot, which are a likely source of scatter in the plot of Ref.~5. Another approach has been to measure the effect of pressure on a single compound. For the case of YBCO, $%
T_{c}$ has been found to initially increase under hydrostatic pressure \cite%
{McElfresh88,Mori91,Koltz91,Sadewasser00}. Under pressure, $J$ also increases\cite{Mallet13}, yielding $\Delta T_{c}^{max}$/$\Delta J$$\sim$1.5 K/meV. While similar order-of-magnitudes are
encouraging, it shows that the fluctuations in the slope could be large
depending on materials or conditions. In fact, Mallet \textit{et al.}\cite{Mallet13}
observed a negative $J$-$T_{c}$ slope in a series of $\mathrm{%
RA_{2}Cu_{3}O_{y}}$ compounds with $A$=(Ba, Sr) $R$=(La,..Lu,Y), casting
doubt on the spin-mediated scenarios.

Another key parameter thought to strongly affect the cuprates is the $%
d_{3z^{2}-r^{2}}\rightarrow d_{x^{2}-y^{2}}$ orbital splitting \cite%
{Ohta91,Pavarini01,Sakakibara10,Kuroki11,Sakakibara12,Yoshizaki12}. This
splitting increases with increasing apical oxygen distance $d_{A}$ from the
copper-oxygen plane. When the splitting grows, it increases the in-plane
character of the holes, creating a condition favorable for superconductivity
by stabilizing the Zhang-Rice singlet \cite{Ohta91} or rounding the Fermi
surface \cite{Sakakibara10}. A higher $d_{A}$ also reduces screening from
polarizeable charge reservoir layers \cite{Raghu12}. All three options are
expected to lead to higher $T_{c}$.

The multitude of different control parameters for $T_{c}^{max}$ emphasizes
the importance of measuring their effects in isolation. Here we measure both
$J$ and orbital splitting in $\mathrm{(Ca_{x}La_{1-x})(Ba_{1.75-x}La_{0.25+x})Cu_{3}O_{y}}$ (CLBLCO), using resonant inelastic x-ray scattering (RIXS).
CLBLCO, whose phase diagram is shown in Fig.~\ref{fig:phase_diagram}, is a compound which allows the tuning of structural parameters
independently of the hole doping. Its structure is almost identical to YBCO\cite{Goldschmidt93},
but it is tetragonal and its chain layers are not ordered. The oxygen
content $y$ controls the number of doped holes, only slightly affecting
the lattice parameter. In complimentary fashion, Ca/Ba content $x$ changes
only structural parameters such as bond length $a$, buckling angles $\theta $, and apical distance $d_{A}$, while keeping the net valence fixed \cite%
{Sanna09}. Additionally, the entire doping range can be spanned from undoped
to overdoped for all values of $x$. Therefore, $x$ tunes both $J$ (through $a$ and $\theta $) and orbital splitting (through $d_{A}$) over the whole
phase diagram. Moreover, disorder in CLBLCO was found to be $x$-independent
based on the line-widths measured by techniques ranging from high resolution powder x-ray diffraction \cite%
{Agrestini14}, Cu, Ca, and O nuclear magnetic resonance \cite{Marchand05,Keren09,Amit10,Cvitanic14}, phonon \cite{Wulferding14}, and ARPES \cite%
{Drachuck14}.

Intriguingly, both $J$, as measured in undoped CLBLCO samples, and $%
T_{c}^{max}$, were found to increase with $x$ by as much as 40\%. In fact,
the energy scale of the entire phase diagrams, including the magnetic, spin glass, and
superconducting parts, scale with $J$ \cite{Ofer06,Ofer08}. Such scaling was
attributed to a superconductivity governed by $J$ \cite%
{Ofer06,Ofer08,Drachuck14,Wulferding14}. However, $J$ in the optimally doped
samples, and the possible effects of the apical oxygen distance, are not
known. Measuring those are the main objective of this work.

Recent milestones in the technique of resonant inelastic (soft) x-ray
scattering (RIXS) have been the measurement of dispersive magnetic
excitations in superconducting cuprates and iron pnictides\cite%
{Braicovich09,Braicovich10,LeTacon11,Zhou13,Dean13}. This, together with
single-crystal CLBLCO growths \cite{Drachuck12}, have allowed us to
measure how $J$ varies with $x$ in both underdoped and optimally doped
CLBLCO samples. The momentum dependence provided by RIXS enables the precise
determination of $J$ based on the spin-wave dispersion. Fortuitously, the
same probe is also sensitive to the orbital $dd$ excitations \cite%
{Ghiringhelli04}. Here we measure of both effects simultaneously on CLBLCO single crystals. We also use x-ray absorption
spectroscopy (XAS) to verify that the effective hole dopings are indeed the
same when we compare families with different $x$. 

We find that $T_{c}^{max}$
has a positive, but not proportional, correlation with $J$. The $d_{xy}\rightarrow d_{x^{2}-y^{2}}$ and $d_{xz/yz}\rightarrow d_{x^{2}-y^{2}}$
splittings also increase with $x$, but we could not precisely isolate the $%
d_{3z^{2}-r^{2}}-d_{x^{2}-y^{2}}$ excitation. We nevertheless determine that in this system $\Delta$$J$ has a greater
contribution (treating it as the independent variable) to the change in $%
T_{c}^{max}$, compared to the out-of-plane orbital effect\cite%
{Ohta91,Pavarini01,Sakakibara10,Kuroki11,Sakakibara12,Yoshizaki12}. We also
observed that the change in $J$ is very similar in the undoped and doped
samples, and the slope of the $J$-$T_c^{max}$ relation is identical to that
of YBCO under pressure. The RIXS spectra also revealed unexpected features,
including a peak at 0.8 eV, and energy dispersive $dd$ excitations.

The paper is organized as follows. Section \ref{sect:experiment} describes
the experimental details. Presentation and analysis of the RIXS data are
divided into the sections: (\ref{sect:RIXSUD}) RIXS of underdoped samples,
with focus on the spin excitations; (\ref{sect:RIXSOD}) RIXS of doped
samples; and (\ref{sect:dd}) $dd$, or ``crystal field'' orbital excitations.
Section \ref{sect:Discussion} is a discussion of the context of these
results, and Section \ref{sect:Conclusion} is the conclusion. Analysis of the doping from the O $K$-edge and Cu $L$-edge x-ray absorption spectra is
provided in the Appendix.


\section{Experimental Details}

\label{sect:experiment}

\begin{figure}[tbp]
\centering
\epsfig{file=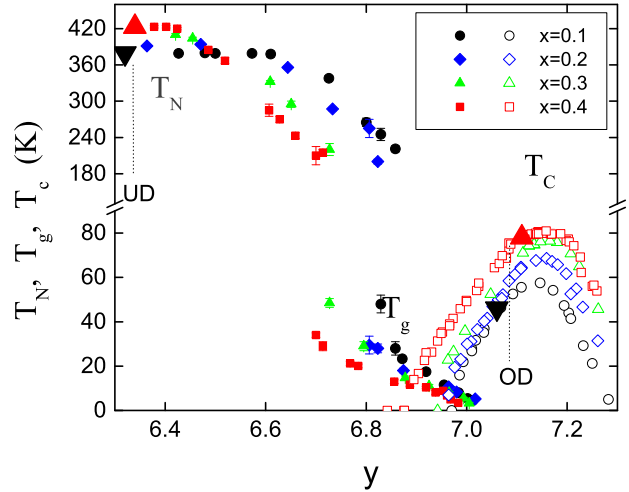,height=2.9in,keepaspectratio}
\caption{(Color Online) Phase diagram of CLBLCO obtained from powder samples in Ref.~ 29, which plots the N\'{e}el ($T_N$), spin-glass ($T_g$), and superconducting ($T_c$) transition temperatures as a function of the stoichiometric oxygen amount $y$ for various families ($x$).  The under-doped (UD), medium doped (MD), and near-optimally doped (OD) single-crystal samples of the present study, are placed in the diagram as triangles, black (pointing down) for $x=0.1$ and red (pointing up) for $x=0.4$.}
\label{fig:phase_diagram}
\end{figure}

($\mathrm{Ca_{x}La_{1-x})(Ba_{1.75-x}La_{0.25+x})Cu_{3}O_{y}}$ single
crystals were grown using the traveling float-zone method \cite{Drachuck12}.
For each of $x$=0.1 and $x$=0.4, the under-doped (UD) samples (in \emph{y}) were prepared by annealing in argon.  The near-optimally doped (OD) samples were first annealed in flowing oxygen, followed by 100 Atm oxygen pressure for a period of two weeks. The oxygen content for the OD samples was confirmed by iodometric
titration.  The oxygen content for UD sample was set based
on the procedures used for powders so as to be in the antiferromagnetically
long-range ordered phase\cite{Ofer06}. The $T_{c}$'s of the $x=0.4$ and $x=0.1$ OD
samples were measured by magnetic susceptibility using a SQUID magnometer.
They were $78$\thinspace K, and $46$\thinspace K respectively, indicating a
close to optimal doping condition. The place of the samples in the phase diagram is shown in Fig.~\ref{fig:phase_diagram}.

Soft x-ray absorption spectroscopy (XAS) and RIXS measurements were conducted at the ADRESS beamline
\cite{Strocov09} at the Swiss Light Source of the Paul Scherrer Institut.
The sample environment was $\sim$10 K in vacuum. The sample surfaces were
cleaved $c$-axis faces, mounted such that the $a$- (or equivalent $b$-) axis
was in the horizontal scattering plane. For XAS, to obtain incident
polarization approximately parallel to the $c$-axis, the sample was rotated
to 10$^{\circ}$ from the grazing incidence condition. Refer to the Appendix for the detailed XAS results. 

RIXS spectra were measured in the horizontal
scattering plane. Measurements were done for both horizontally and
vertically polarized incident beams, corresponding to $\pi$ and $\sigma$
polarizations respectively. The incident energy was set to the first main
peak in the Cu $L_{III}$ XAS at 932 eV.
The detector was fixed such that the two-theta scattering angle with respect
to the incident beam was 130$^{\circ}$. Throughout this article, we refer to the in-plane momentum
transfer $q$ in reciprocal lattice units of 2$\pi$/$a$, where $a$ is the
lattice constant of the crystal.  We define $q$ as the component of the total momentum change of the photon which is parallel to the sample $ab$ plane \cite{Braicovich09}.  Our sign convention is grazing incidence corresponding to negative $q$.  The variation of $a$ with each $x$\cite{Ofer08} is accounted for in calculating $q$, but is not significant on the $q$-scale. In our
scattering configuration $q$ is always along the (1 0 0) direction, and its
magnitude is changed by rotating the sample away from specular reflection.
Therefore the total momentum transfer is $\textbf{Q}$=($q$, 0, $L$) in tetragonal notation.  The grazing incidence condition
was used to calibrate the $q$ position. This calibration was found to be valid by measuring $E$ vs. $q$ dispersion for both positive and negative $q$ (see Section~\ref{sect:RIXSUD}).

\section{Magnons in Underdoped Samples}

\label{sect:RIXSUD}

\begin{figure}[tbp]
\centering
\epsfig{file=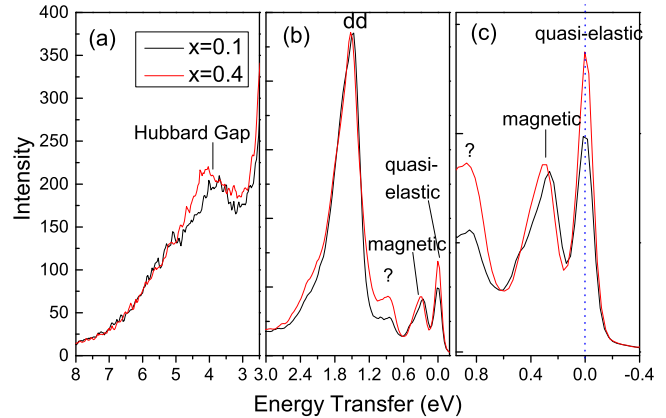,height=2.4in,keepaspectratio}
\caption{(Color Online) The main features of the typical CLBLCO RIXS spectra
for $x$=0.1 UD (black) and $x$=0.4 UD (red) at various energy ranges: (a)
3-8 eV (b) 0-3 eV and (c) below 1 eV. In this example, the incident beam is $%
\protect\pi$-polarized, and $q$=-0.34}
\label{fig:RIXS_UD}
\end{figure}

Typical RIXS spectra for the UD samples are compared in Fig.~\ref{fig:RIXS_UD} for $x=0.1$ and $x=0.4$. The
various panels of Fig.~\ref{fig:RIXS_UD} zoom in on different energy scales.
The intensities are normalized to match at the strong $dd$ peak in Fig.~\ref%
{fig:RIXS_UD}(b), and the energies are shifted so that the
quasielastic peaks in Fig.~\ref{fig:RIXS_UD}(c) are centered at zero.
Fig.~\ref{fig:RIXS_UD}(a) shows the relatively high-energy part of the spectra. First, we note that the $x=0.1$
and $x=0.4$ tails going down from 5 eV overlap closely. Secondly, there is
a peak at around 4.5 eV which is in the energy range of charge transfer
excitations across the Hubbard gap. The feature is shifted to higher energy
for $x=0.4$, and is also present in the doped samples. A doping-independent
feature at similar energy was studied in LSCO with Cu $K$-edge RIXS \cite%
{Ellis11}. Fig.~\ref{fig:RIXS_UD}(b) is an overall view of the spectra
including both the intense peak encompassing the $dd$ excitations between $%
1.5$ and $1.8$ eV, and the lower energy peaks, which are much lower
intensity, but still visible on this scale. Comparison of the $x=0.4$ and $%
x=0.1$ spectra over a broad range reveals that generally, most of the
excitations are at slightly higher energy for $x=0.4$.  This increased energy is ubiquitous both for the magnon excitations covered in this and the following section, and for the $dd$ excitations in Section~\ref{sect:dd}.  We note that both the high energy tails in Fig.~\ref{fig:RIXS_UD}(a), and the quasielastic peaks in Fig.~\ref{fig:RIXS_UD}(c), are aligned in energy for the two samples.  We will later show that the magnon and $dd$ excitation energy increases can be directly attributed to the change in lattice parameters.

Fig.~\ref{fig:RIXS_UD}(c) zooms in on the low energy range. At zero energy
is a quasielastic peak, which depends on a combination of finite $q$%
-resolution of the instrument and the crystal mosaic of the sample. In
our analysis, the energy scale of each spectra is shifted according to the
center energy of the quasielastic peaks. In similar measurements done by
Braicovich \textit{et al.} \cite{Braicovich10} for $\mathrm{La_2CuO_4}$, in which the
quasielastic peak was much lower than it is here, a feature at around 80-90 meV
was observed, with about a fifth of the magnon intensity. This was
attributed to a resonantly enhanced optical phonon. We did not detect such a phonon and it is not included it in
our analysis.

\begin{figure}[htbp]
\centering
\epsfig{file=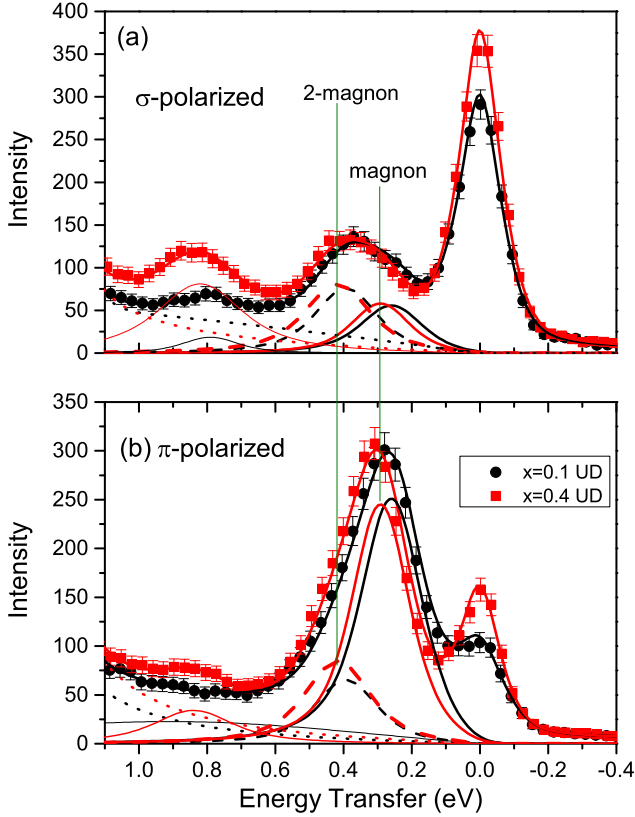,height=4.3in,keepaspectratio}
\caption{(Color Online) The RIXS spectra in underdoped samples at $q$=0.375
for (a) $\protect\sigma$ and (b) $\protect\pi$ polarized configurations,
corresponding to vertical and horizontal polarizations of the incident beam
for scattering in the horizontal plane. The spectra of $x$=0.1 (black) and $%
x $=0.4 (red) samples are compared. The magnon ($\sim$0.25-0.3 eV) and
2-magnon ($\sim$0.4 eV) component peaks are indicated as solid and dashed
lines, obtained from simultaneous fitting of both (a) and (b) spectra. The
dotted line shows a fit of the tail of the higher energy $dd$ excitation. A
peak around at 0.8 eV, much stronger in the $x$=0.4 sample, is also present.
The total fits are shown as solid lines crossing the data.}
\label{fig:RIXSfit}
\end{figure}

The peak associated with magnons is found in the 0.2-0.4 eV range of Fig.~%
\ref{fig:RIXS_UD}(c). Comparison of the data for $x=0.4$ (red) and $x=0.1$
(black) clearly shows that the $x=0.4$ peak is shifted to higher energy. Thus the main result that $J$ is higher in the $x=0.4$ sample than in the $x=0.1$ samples is clearly evident already in the raw data.

There is also a peak at $\sim 0.8$ eV. Its intensity is highest (comparable
to the magnetic peak) at negative $q$ for $\pi $ polarized scattering, but
can be seen elsewhere (see Fig.~\ref{fig:RIXSfit}) and is always stronger
for the $x=0.4$ sample. Where it is large, it was incorporated into our
fitting for the magnons, described below. It has only slight dispersion of $<0.05$ eV, unlike the new mode recently observed by Lee \textit{et al.}\cite{Lee14}.
It would be surprising if the $\sim 0.8$ eV peaks were one of the three $dd$
excitations, which are expected to be above 1.5 eV\cite{Sala11}. On the
other hand, a $dd$ excitation in the chain layer would be more plausible.
Since half of the non-apical oxygen ligands around each Cu atom are missing
in the chain, the Coulomb energy cost for a chain $dd$ excitation should
also be about half of a plane $dd$ excitation, which corresponds to this $%
\sim 0.8$ eV.

\begin{figure}[htbp]
\centering
\epsfig{file=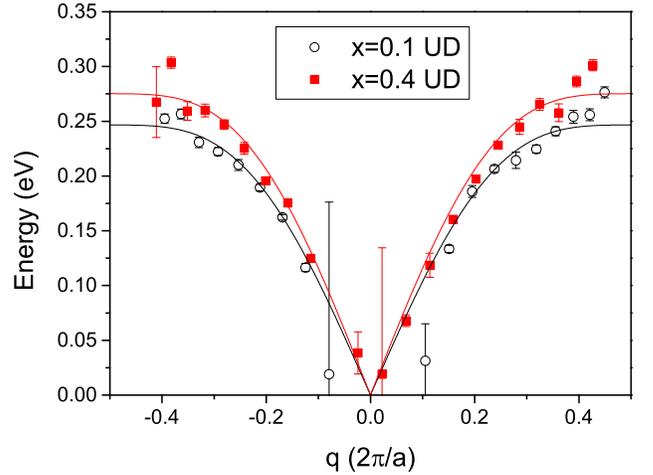,height=2.5in,keepaspectratio}
\caption{(Color Online) The dispersion along the (1 0 0) direction, of the
fitted energies $\protect\nu_R$ of the single-magnon components of the $x$%
=0.1 (black) and $x$=0.4 (red) UD samples. Fits to theoretical acoustic
magnon dispersions of Ref.~43 for free parameter $J$,
with $J_{\perp }$ fixed to 15 meV, are shown as black ($x=0.1$) and red ($%
x=0.4$) lines.}
\label{fig:magdispUD}
\end{figure}

A sample pair of spectra corresponding to the $\sigma$ and $\pi$
polarizations at the same $q$ are shown in Fig.~\ref{fig:RIXSfit}(a) and
(b). To extract the magnon energies, fitting was done over the range
shown in Fig.~\ref{fig:RIXSfit}. Each spectrum was modeled as a sum of
quasielastic peak, magnon, 2-magnon, with (when visible) an additional peak
at 0.8 eV, and a tail from the $dd$ excitations. The spectral weight of the
2-magnon relative to the magnon is generally different for the $\sigma$ and $%
\pi$ polarizations, resulting in a shift in the peak energy for the
different polarizations. As in Ref.~33, the fitting is done for both
polarizations simultaneously. The energies and widths of the magnon and
2-magnon peaks were constrained to be the same for both polarizations, as
indicated by the vertical lines in Fig.~\ref{fig:RIXSfit}. The lineshapes as
a function of energy $\nu$ used for all of the excitations was a damped harmonic oscillator response in the form of a
Lorentzian, weighted according to ``detailed balance'' :

\begin{align}
S(\nu )=
\begin{split} 
&\frac{1}{1-e^{-\nu /k_{B}T}} \\
&\times\bigg(\frac{(\Gamma /2)^{2}}{(\nu -\nu_{R})^{2}+(\Gamma /2)^{2}}-\frac{(\Gamma /2)^{2}}{(\nu +\nu_{R})^{2}+(\Gamma /2)^{2}}\bigg)   \label{eqn: lineshape}  
\end{split}
\end{align}%
where $T$ is the sample temperature, $k_{B}$ is Boltzmann's constant and fit
parameters $\nu _{R}$ and $\Gamma $ are the energy and intrinsic width
respectively. Each $S(\nu )$ was then convolved with a Gaussian representing
the resolution function of the spectrometer, to produce the components shown
in Fig.~\ref{fig:RIXSfit}. The fits for all spectra (more than 80) were
excellent and are shown in the Supplementary materials \cite{SupMat}. The dispersion of $%
\nu _{R}$ for the magnon components is plotted in Fig~\ref{fig:magdispUD}.
The horizontal $q$-axis for each sample was corrected by a slight shift
(0.013 for $x$=0.1 and 0.022 for $x$=0.4) to make each dispersion
symmetrical about the origin.

The dispersions are fit to a theoretical expression for acoustic-mode
dispersion in the double layer cuprate YBCO \cite{Hayden96}:

\begin{equation}
E=2J\cdot\big(\;1-\gamma^{2}(q)+(\,J_{\bot }/2J\,)\cdot(\,1-\gamma(q)\,)\;\big)^{2}  \label{eqn: dispersion}
\end{equation}%
for in-plane magnetic exchange $J$, with interplane coupling $J_{\bot }$ set to 15 meV, and $%
\gamma(q) =0.5\cdot (cos(2\pi \cdot q)+1)$. There is also in principle an
optical mode\cite{Hayden96}, but it resides quite close to the acoustic
mode, except at low $q$, where the errorbars are high. In our fitting, $%
J_{\bot }$ was fixed at 15 meV (which is similar to YBCO \cite{Hayden96}),
so the only free parameter was $J$. The fits are shown as the lines in Fig.~%
\ref{fig:magdispUD}. Eq.~\ref{eqn: dispersion} captures the non-linearity of
the data, particularly well on the negative $q$ side. The resultant $J$
values were $134\pm 1$ meV for $x=0.4$ and $120\pm 1$ for $x=0.1$.

These values should be compared with detailed $ab$ $initio$ calculations
done by Petit and Lepetit for optimally doped CLBCO\cite{Petit09}. Those
yielded mean values of $J=132$ meV for $x=0.4$ and $J=110$ meV for $x=0.1$.
The $x=0.4$ results are in excellent agreement between theory and
experiment, while there is a $10$ meV difference for $x=0.1$.  We show in
the next section that the dispersion of the UD and OD samples are similar.

\begin{figure}[tbph]
\centering
\epsfig{file=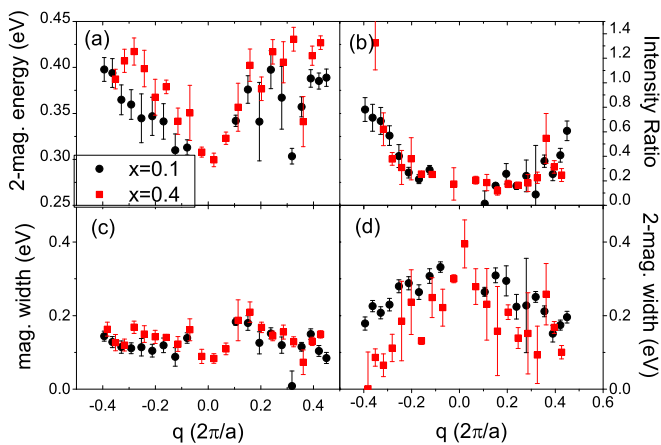,height=2.3in,keepaspectratio}
\caption{(Color Online) 
The other fit parameters for the $x$=0.1 UD (black) and $x$=0.4 UD (red) samples: (a) 2-magnon energy (b) ratio of 2-magnon to magnon intensities in the $\protect\pi $-polarized spectra (c) intrinsic magnon width (FWHM) and (d) intrinsic 2-magnon width.}
\label{fig:otherparams}
\end{figure}

The other fit parameters are plotted in Fig.~\ref{fig:otherparams}(a)-(d).
In Fig.~\ref{fig:otherparams}(a), the 2-magnon energies at low $q$ are close
to $0.3$ eV.
This magnitude is within the range of the recent 2-magnon Raman study in this material by
Wulferding \textit{et al.} \cite{Wulferding14}, who measured energies of ~$0.29-0.35$~eV in various samples. In addition, the sign
and magnitude of dispersion of the 2-magnon of about ~$0.1$ eV in Fig.~\ref%
{fig:otherparams}(a) is reasonably consistent with the $\sim $80 meV
measured with O $K$-edge RIXS by Bisogni \textit{et al.}\cite{Bisogni12} in $\mathrm{%
La_{2}CuO_{4}}$ (see Figure 6 of Ref.~44). Fig.~\ref{fig:otherparams}(b)
plots the ratio of the intensities of the 2-magnon to the 1-magnon
components. They fall on the same curve for $x=0.1$ and $x=0.4$, which is
expected since the excitations in both should have the same symmetries.

Fig.~\ref{fig:otherparams}(c) and (d) show the intrinsic widths $\Gamma $
for the magnon and 2-magnon, which are 100-150 meV and $>300$ meV
respectively. These are wider than expected. The 2-magnon widths observed in
the Raman study~\cite{Wulferding14} were only $\sim $100 meV, while the
magnon width is expected to be resolution limited on this scale. It is not
clear if the large width originates in the fitting or sample.  There is some intrinsic disorder in the site occupation between Ca, Ba, and La atoms, which could be a potential cause of an intrinsic magnon width.  But if so, we note that the widths of $x$=0.1 and $x$=0.4 are about the same, indicating that $x$ does not affect disorder. Nevertheless,
our analysis: (1) fit all of the data excellently with minimal number of
parameters, (2) resulted in a realistic dispersion curve with $J$ values
which are in good agreement with Ref.~45, and (3) 2-magnon energies at low $q$ are consistent with the 2-magnon energies measured with Raman scattering \cite{Wulferding14}, and (4) 2-magnon dispersion is consistent with O $K$-edge value for $\mathrm{La_{2}CuO_{4}}$ from Ref.~44. 

\section{Paramagnons of Optimally Doped Samples}

\label{sect:RIXSOD}

\begin{figure}[htbp]
\centering
\epsfig{file=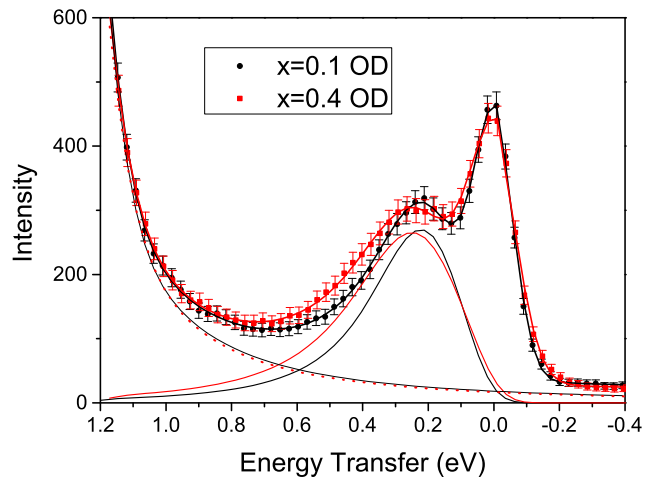,height=2.5in,keepaspectratio}
\caption{(Color Online) $\protect\pi$-polarized spectra of $x$=0.1 (black)
and $x$=0.4 (red) OD samples, at q=0.266. The fit of the magnetic component
to an asymmetric Lorentzian plus background, and the magnetic peak component
itself, are shown as lines with the same color code. The high-energy $dd$
tail of $x$=0.4 closely coincides with that of $x$=0.1, and is shown as a
dotted line.}
\label{fig:ODspec}
\end{figure}

Here we estimate the change in $J$ in the superconducting samples. For doped
cuprates, Le Tacon \emph{et al.} found that the lifetime broadening of the
spin excitations make the widths too broad to distinguish between magnon and
2-magnon, and instead they are replaced by a single \textquotedblleft
paramagnon\textquotedblright\ peak \cite{LeTacon11}. A typical spectrum for
the OD CLBLCO samples is shown in Fig.~\ref{fig:ODspec}. As in Ref.~33 we
replaced the magnon and 2-magnon with a single magnetic component, retaining
the lineshape of Eq.~\ref{eqn: lineshape}. Only the elastic intensity,
\textquotedblleft paramagnon\textquotedblright\ peak, and tail from the $dd$
were included in the fits. Most of the $q$'s measured were positive, and
there were no strong 0.8 eV peaks. Since the peak position is generally
different for $\pi $ and $\sigma$ polarizations, due to different weights of the 2-magnon and magnon contributions (as seen in Fig.~\ref{fig:RIXSfit}), both could not be fit
simultaneously with one peak.  We therefore chose to use only the $\pi $
polarization. The single peak of Eq.~\ref{eqn: lineshape} plus background
fit quite well to the data; fits to all of the spectra are shown in the
Supplementary materials \cite{SupMat}. As seen in both the fits and raw data of Fig.~\ref%
{fig:ODspec}, the paramagnon for $x=0.4$ is shifted with respect to $x=0.1$
and extends to higher energy, which was also typical for the other $q$%
's. We note that in Fig.~\ref{fig:ODspec} the $dd$ tails from high energy
are the same for $x=0.1$ and $x=0.4$.

\begin{figure}[htbp]
\centering
\epsfig{file=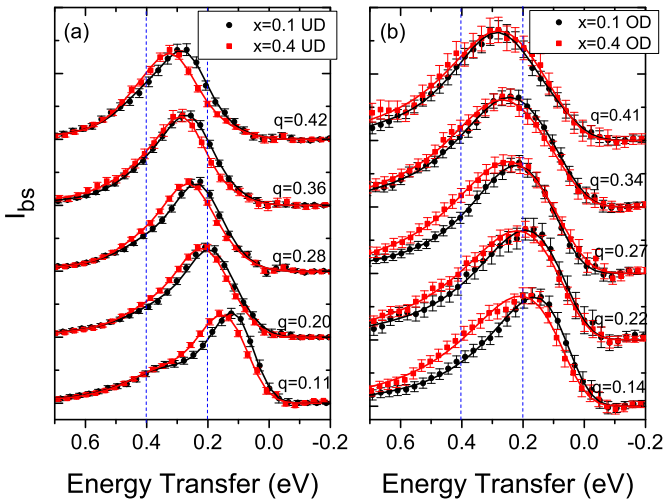,height=2.6in,keepaspectratio}
\caption{(Color Online) Comparison of the background-subtracted spectral intensity ($I_{bs}$)
between (a) UD and (b) OD samples measured at similar $q$ positions. The
spectra were obtained by subtracting the quasielastic, $dd$, and 0.8 eV
fitted components (if present) from the $\protect\pi$-polarized spectra. The
dashed vertical lines are guides for the eye.}
\label{fig:UD_OD_comparison}
\end{figure}

A series of spectra for progressively higher $q$ are plotted in Fig.~\ref%
{fig:UD_OD_comparison}, for (a) the UD samples and (b) the OD samples. The
spectra in Fig.~\ref{fig:UD_OD_comparison}(a) for the UD samples were
obtained by subtracting all of the fitted components (see Section~\ref%
{sect:RIXSUD}) from the raw data, save for the magnon and 2-magnon
contributions. The same procedure is applied to the OD samples in Fig.~\ref{fig:UD_OD_comparison}(b), by subtracting the non-magnetic contribution.
The $q$ positions are similar for Fig.~\ref{fig:UD_OD_comparison}(a)
and Fig.~\ref{fig:UD_OD_comparison}(b). Both pairs of spectra in Fig.~\ref%
{fig:UD_OD_comparison}(a) and (b) are centered below 0.2~eV at low $q$ (bottom spectra), and by $q$=0.4
(top spectra) they dispersed to 0.3 eV. This similarity suggests that the $J$
comparison for the UD spectra, which is generally easier to precisely
determine, is also valid for the superconducting case. It also would seem to argue
against the scenario of intraband excitations (as opposed to paramagnons)
which was recently proposed by Benjamin et al.\cite{Benjamin14}, since the
OD and UD spectra have the same energies. 

The value of $J$ cannot directly be determined from the \textquotedblleft
paramagnon\textquotedblright\ spectra. The fitted energy parameter $\nu _{R}$
of the asymmetric lineshape in Eq.~\ref{eqn: lineshape} does not have the
same well-defined meaning as in the two-peak, two-polarization fits used in
section~\ref{sect:RIXSUD}. This is because the peak fitted-for here
encompasses both magnon and 2-magnon components, weighted by some unknown
amount depending on scattering cross-section for each (one can refer to Fig.~%
\ref{fig:otherparams}(b) for the UD case). Instead, for comparison purposes
we use the center-of-mass, namely, the statistical mean energy $\left\langle
E_{M}\right\rangle $=$\frac{\int E\cdot I_{bs}(E)dE}{\int I_{bs}(E)dE}$ of
the background-subtracted magnetic spectra $I_{bs}(E)$ of Fig.~\ref%
{fig:UD_OD_comparison}(b). While this definition is arbitrary, for a given $q$, it should be
roughly proportional to $J$ for any two samples, since both
magnon and 2-magnon energies are proportional to $J$.

\begin{figure}[htbp]
\centering
\epsfig{file=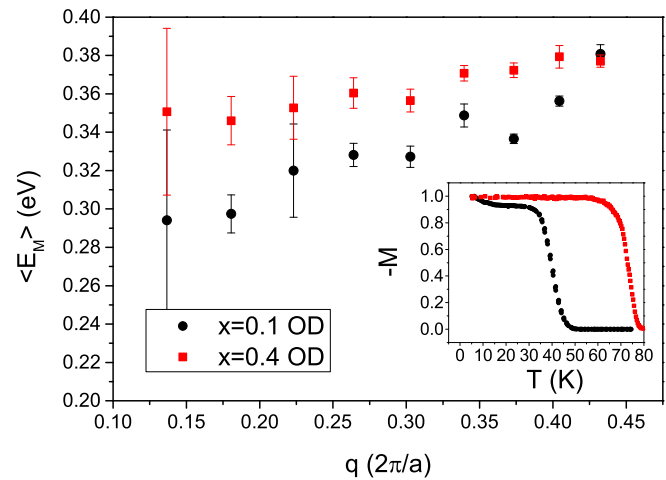,height=2.5in,keepaspectratio}
\caption{(Color Online) The dispersion of the energy center-of-mass $%
\left\langle E_{M}\right\rangle$ of the magnetic peak of the OD samples as
described in the text. To emphasize their $T_c$ variations, the inset shows the magnetization versus temperature of the two samples, normalized to their maximum diamagnetic responses at low temperature.}
\label{fig:COMdisp}
\end{figure}

$\left\langle E_{M}\right\rangle $ is plotted as a function of $q$ in Fig.~%
\ref{fig:COMdisp} for $x=0.1$ (black circles) and $x=0.4$ (red squares). For
all but the last, it is higher for $x=0.4$. The average over these $q$
points, $\overline{\!\left\langle E_{m}\right\rangle }$, is $0.33$ eV for $%
x=0.1$ and $0.36$ eV for $x=0.4$. Assuming proportionality, we interpret
this as a 9\% increase in $J$ from $x=0.1$ to $x=0.4$. By comparison, the
percentage increase for the (more precisely determined) $J$'s of the UD
samples in Section~\ref{sect:RIXSUD} is 11.7\%. Considering the broad widths
of the OD spectra and the somewhat cruder method of estimating their $\Delta
J$, this estimated increase is quite close to the UD case.

In the inset of Fig.~\ref{fig:COMdisp} we present the negative magnetization measurements of the two superconducting samples used for RIXS. There is a clear difference in their $T_c$. The main observation of this work is that the sample with higher $T_c$ also has higher $J$.  It was also demonstrated here that RIXS can distinguish samples with small differences in $J$ even in the optimally doped case.

\section{Crystal Field ($dd$) Excitations}

\label{sect:dd}

\begin{figure}[htbp]
\centering
\epsfig{file=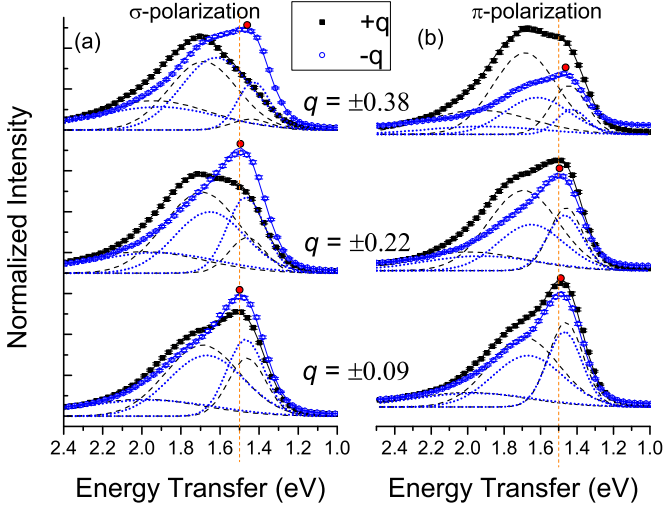,height=2.7in,keepaspectratio}
\caption{(Color Online) $dd$ spectra of the UD sample for $x$=0.1 at representative $q$ positions for (a) $\sigma$-polarization and (b) $\pi$ polarization.  The three Gaussian components are
indicated as dashed/dotted lines, and the total fit as solid lines.  The filled black squares/dashed lines correspond to positive $q$, the empty blue circles/dotted lines correspond to negative $q$ (closer to grazing incidence).  The vertical lines and red circles are guides for the eye.}
\label{fig:dd_spectra_and_fits}
\end{figure}

The $dd$ spectra of our UD samples were generally sharper than for our OD
samples, so we focus on the former. The $dd$ excitation spectra of the UD
samples are plotted in Fig.~\ref{fig:dd_spectra_and_fits} for selected $q$%
's for the $x$=0.1 sample.  The spectra of the $x$=0.4 sample was qualitatively similar in the main features, but with slightly higher energies (see Fig.~\ref{fig:RIXS_UD}(b)). All of the spectra and fittings for the full range of $q$'s are presented in the Supplementary materials\cite{SupMat}.  The centering of the quasi-elastic peaks of all of the spectra are also shown in the Supplementary materials to be accurate within $\sim$10 meV.  At least two peaks are clearly resolved, at $\sim$1.5 eV and $\sim$1.7
eV, with the intensity of the 1.7~eV peak becoming relatively stronger with
increased $q$.  We fit the $\pi$ and $\sigma$ polarized spectra simultaneously
to a sum of Gaussians, constraining the parameters of widths and energies to
be the same for both polarizations. Three Gaussians worked best. They are shown in
Fig.~\ref{fig:dd_spectra_and_fits}. The zero energies are defined by the elastic
peaks (see Section~\ref{sect:RIXSUD}). As can be seen in Fig.~\ref{fig:dd_spectra_and_fits} the widths successively increased from the low to high energy peaks.

To assign the peaks, we refer to the work of Sala \textit{et al.}\cite{Sala11}, who
studied $dd$ excitations with Cu $L$-edge RIXS in a variety of cuprates.
They found excellent agreement between the observed polarization and $q$
dependence, and their cross-section calculations. The compound studied in
that work which is structurally similar to CLBLCO is the double-layer
123-cuprate $\mathrm{NdBa_2Cu_3O_7}$ (NBCO). In what follows, $E_{xy}$, $%
E_{xz/yz}$, and $E_{3z^2-r^2}$ refer to the energies of the orbital transitions $%
d_{xy}\rightarrow d_{x^{2}-y^{2}}$, $d_{xz/yz}\rightarrow d_{x^{2}-y^{2}}$,
and $d_{3z^2-r^2}\rightarrow d_{x^{2}-y^{2}}$ respectively. The
NBCO spectra had two prominent peaks at 1.52~eV and 1.75~eV, which the
authors of Ref.~42 assigned to $E_{xy}$ and $E_{xz/yz}$. $E_{3z^2-r^2}$ was
calculated to be 1.97~eV, but it was not visible in their spectra. As $q$
increased, the cross-section of the 1.75~eV peak increased relative to the
1.5~eV peak. These results, both the energies and cross-section $q$%
-dependence are very close to what we observe for CLBLCO in Fig.~\ref%
{fig:dd_spectra_and_fits}. We therefore likewise assign the 1.5 eV peak to $%
E_{xy}$ and the 1.7 eV peak to $E_{xz/yz}$. Furthermore, the energy of the
broad third Gaussian component in Fig.~\ref{fig:dd_spectra_and_fits}
happened to lie very close to 2 eV, with zone-averages (standard deviations)
of 1.97(0.03) eV, and 2.00(0.1) eV, for $x$=0.1 and $x$=0.4 respectively.
While this energy is in excellent agreement with calculations for $%
E_{3z^2-r^2}$ in NBCO\cite{Sala11}, and for YBCO\cite{Magnuson14} the
broadness makes it difficult to identify with certainty.

\begin{figure}[htbp]
\centering
\epsfig{file=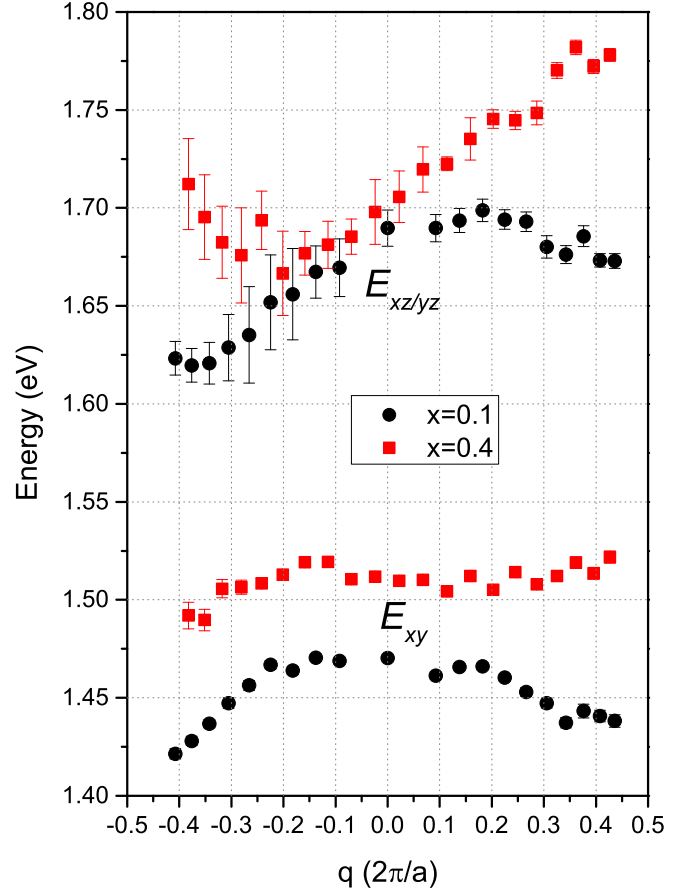,height=4.7in,keepaspectratio}
\caption{(Color Online) The fitted $dd$ energies $E_{xy}$ and $E_{xz/yz}$ which resulted from the simultaneous fitting to both $\pi$ and $\sigma$ polarizations, plotted for each $q$ position. The energies of $x$=0.1 (black) and $x$=0.4
(red) are plotted.}
\label{fig:dd_energies}
\end{figure}

The $q$-dependence of $E_{xy}$ and $E_{xz/yz}$ are plotted in Fig.~\ref%
{fig:dd_energies} for $x$=0.1 and $x$=0.4. Surprisingly, there appears to be
some dispersion in the energies.  The $d_{xy}$ excitation for $x$=0.1 shows flat dispersion near the zone center, up to around $|q|$=0.2, but beyond this it exhibits negative dispersion of
the order of 0.05 eV towards the zone boundary. This $E_{xy}$ dispersion is quite symmetrical about $q$%
=0, up to $q$=0.35.  The dispersion could also be seen from the raw data.  In Fig.~\ref{fig:dd_spectra_and_fits}, the red circles mark the low-energy peaks of the negative $q$ branch. They also mark the positive branch, but it is harder to see for high $|q|$, especially for $\sigma$-polarization. For the bottom two spectra, corresponding to $|q|$=0.09 and $|q|$=0.22, the peaks are aligned with the vertical dashed line, but by $|q|$=0.38, the peak of the raw data is visibly shifted to the right of the line by about 50 meV. A non-zero dispersion suggests propagation of the orbital excitation.  For $x=0.4$, the $d_{xy}$ excitation shows similar dispersion on the negative branch, but its magnitude is roughly halved. $E_{xz/yz}$ also shows dispersion, but is not
symmetrical about $q$=0.  The fitted energies for $x$=0.4 especially show a linear trend with a dispersion of almost 0.1 eV  between $q=\pm0.2$.  Unlike the $E_{xy}$ dispersion, the $E_{xz/yz}$ dispersion is not obvious from the raw data itself due to the wider peaks, and only becomes apparent after the fittings.  Although it seems counter-intuitive, asymmetric dispersion may happen in the presence of spin-orbit interaction.  It has already been observed in the spin-wave of Fe ultrathin films\cite{Zakeri10}, for example. But, as far as we know this would be the first observation of asymmetry in the dispersion of a $dd$ excitation.

We can check whether $E_{xy}$ scales properly with the lattice parameter. As
pointed out by Sala et al.\cite{Sala11} $E_{xy} \propto a^{-n}$. Averaging the energies of Fig.~\ref%
{fig:dd_energies} over the zone yields, for $x=0.1$ ($x=0.4$), $\overline{E}%
_{xy}=$1.46 (1.52) eV and $\overline{E}_{xz/yz}=$1.69 (1.75) eV. The corresponding $a$ values for $x=0.1$ is $a$=3.91 \AA\ and for $x=0.4$ is $a$=3.88 \AA. This yields $n$=5.1 remarkably close to
the theoretical single-ion crystal field model's value of $n$=5.

\section{Discussion}

\label{sect:Discussion}

Analysis of the UD spectra in Section~\ref{sect:RIXSUD} provided explicit $J$
values of 120 meV ($x$=0.1) and 134 meV ($x$=0.4).  The corresponding $T_{c}^{max}$ for these $x$ values
are 57 K and 80 K respectively \cite{Ofer06}. In section~\ref%
{sect:RIXSOD}, we found that the change in $J$ for doped samples is
comparable to the undoped case, and the two dopings furthermore exhibit very
similar dispersions of the spin-excitation spectra (refer to Fig.~\ref%
{fig:UD_OD_comparison}). It is therefore justified to apply the UD values of
$J$ to the superconducting case, as has been assumed to be valid in other
works\cite{Mallet13,Wulferding14}. With $x$ as an implicit parameter we find
that $\Delta T_{c}^{max}/\Delta J$=1.64 K/meV. This is the same order of
magnitude of the average slope obtained from the study of Munoz et al.\cite%
{Munoz00} of several cuprates having different numbers of layers ($\sim 3.2$
K/meV) . It is even more closely aligned with the initial slope for YBCO
under hydrostatic pressure ($\sim 1.5$ K/meV) \cite{Mallet13}. 
Moreover, the increase of $J$ of 11.7\% from $x$=0.1 to $x$=0.4 determined for the UD samples in Section~%
\ref{sect:RIXSUD} is in close agreement with the estimation of 11.9\% we obtain by using a simple $%
J\varpropto \cos^{2}\theta /a^{14}$ rule\cite{Ofer08}. 
In addition, $E_{xy}$ scales as expected with distances. 
These results indicate that the
in-plane energies $J(x)$ and $E_{xy}(x)$ depend purely on in-plane parameters,
without secondary effects arising from different Ca/Ba ratios. We speculate that the $d_{3z^2-r^2}\rightarrow d_{x^{2}-y^{2}}$ peak, which we could not properly resolve, behaves as expected from the lattice parameters variations between different CLBLCO families.

Whether $T_{c}^{max}(x)$ likewise depends only on the in-plane parameters is not \textit{a priori} clear, since the out-of-plane lattice parameter $c$, and apical oxygen distance $d_A$ are also functions of $x$. In fact, a number of studies\cite{Ohta91,Pavarini01,Sakakibara10,Kuroki11,Sakakibara12,Yoshizaki12,Johnston10} focused on the effect of $d_A$ and $E_{3z^2-r^2}$ on $T_{c}^{max}$ in various cuprate systems.  We now assess the relative importance that these have for $T_{c}^{max}(x)$.

Since YBCO and CLBCO share very similar structure and lattice parameters, it
is relevant to compare the two. The values of $\Delta T_{c}^{max}/\Delta J$ observed in the pressure-dependence of YBCO on one hand, and in the $x$-dependence of CLBLCO observed here on the other, are very similar. Hydrostatic pressure compresses the
$c$-axis, decreasing the apical oxygen distance $d_A$ and increasing $T_c$. 
In contrast, when increasing $x$ (and $T_{c}^{max}$) in CLBLCO, $d_A$ increases \cite{Ofer08}. 
That $\Delta T_{c}^{max}/\Delta J$ is the same for YBCO and CLBLCO, in spite of $d_A$ changing in the opposite sense, leads us to conclude that $d_A$ variations do not play a major role here in determining $T_c^{max}$. 

Another way to reach this conclusion for CLBLCO is to estimate the effect of the change in $d_A$ on $T_{c}^{max}$ by comparing with other studies. A sensitivity of roughly $\frac{%
	\partial T_{c}^{max}}{\partial d_{A}}\sim $30 K/$\mathrm{\mathring{A}{}}$,
was shown across various cuprates by Johnston et al.\cite{Johnston10} (see
Figure 1 of Ref.~49). In CLBLCO powder,
as $x$ increases from 0.1 to 0.4, $d_A$ increases by $\sim$0.05 $%
\mathrm{\mathring{A}{}}$ \cite{Ofer08}.  On that basis, the effect of $\Delta d_A$ on $T_{c}$
in CLBLCO would be less than $2$ K. 

A similar effect of $\Delta d_A$ on $T_{c}$ results from the theoretical calculations of $%
E_{3z^2-r^2}$($d_A$) by Sakakibara \textit{et al.}\cite%
{Sakakibara10}. They calculated the Eliashberg eigenvalue $\lambda$ which sets a limit on $T_c$. From their calculations, an upper limit of  $\frac{%
\partial T_{c}^{max}}{\partial d_{A}} < $125 K/$\mathrm{\mathring{A}{}}$ can be set, which is still too small to account for the $T_c$ variations in CLBLCO.


Taken together, the above comparisons suggest that $\Delta d_A$ in CLBLCO has very little impact on $T_{c}^{max}$. By eliminating this out-of-plane influence, it becomes more likely that the change in $T_{c}^{max}$ observed between different families of CLBLCO is due to variations in $J$. While $T_{c}^{max}$ increases
by 40\% (Fig.~1) from $x$=0.1 to $x$=0.4, $J$ as measured by RIXS only increases by $\sim 11.7$\%. 
From other methods, the corresponding increase in $J$ for samples with the same in-plane hole underdoping was determined to be: $21\%$ from the 2-magnon Raman peaks \cite{Wulferding14}, $26\%$ from angle resolved photoemission spectroscopy\cite{Drachuck14b}, $20\%$ in \textit{ab initio} calculations \cite{Petit09}, and 40\% by $\mu $SR with extraction of $J$ from $T_N$ \cite{Ofer06}. With the exception of the latter, these estimates were all
considerably less than the increase in $T_{c}^{max}$. 
This suggests that the $J$ dependence of $T_c$  is not proportional, as predicted by some exchange-driven theories\cite{Scalapino98,Sushkov04}. If a linear relationship extends down to $T_{c}^{max}$=0, it would imply a threshold $J$ for superconductivity.

\section{Conclusion}

\label{sect:Conclusion} To review, we measured the O $K$-edge and Cu $L$-edge XAS, and RIXS spectra at the Cu $L$-edge, in both underdoped and optimally doped CLBLCO single crystals of $x=0.1$ and $x=0.4$ families which have different $T_c^{max}$.
 
From the electronic structure of the XAS spectra, similar hole dopings in the superconducting samples of the different families were confirmed.  As it turns out, doping does not have a critical effect on the magnon dispersion, besides a broadening of the peaks.  The relative change in magnetic energies between $x$=0.1 and $x$=0.4 are furthermore similar for the doped and undoped cases.  This demonstrates that RIXS can distinguish between samples of slightly different $J$ even in the doped case.  

The main $dd$ excitations were also examined and unexpectedly
dispersion of up to ~0.05 eV was observed, raising the possibility that these orbital excitations can propagate. More intriguingly, the dispersion of the excitation from the $d_{xz/yz}$ orbit appeared to be asymmetric about $q$=0. Higher resolution studies would be needed to clarify this dispersion. In the UD samples, an additional 0.8 eV peak was observed, and attributed to a $dd$ excitation in the chain layer. 

Finally, there is a positive
correlation between $T_c^{max}$ and $J$ with a slope consistent with the
pressure dependence of both parameters in YBCO. The measured spin-wave energies change with $x$ by an amount that would be
expected from purely in-plane lattice constants change. Furthermore, it is concluded that the apical oxygen distance does not change enough with $x$ to have a significant effect on $T_c^{max}$. These points suggest that the $T_c^{max}$ variation with $x$ in CLBLCO is purely an in-plane effect driven by orbital overlaps.

\section*{Acknowledgements}

The RIXS and XAS measurements were performed on the ADRESS beamline at the
Swiss Light Source, Paul Scherrer Institut, Villigen, Switzerland. Part of this work has been funded by the Swiss Nationional Science Foundation and its Sinergia network Mott Physics Beyond the Heisenberg (MPBH) model. J.P. and T.S. acknowledge financial support through the Dysenos AG by Kabelwerke Brugg AG Holding, Fachhochschule Nordwestschweiz, and the Paul Scherrer Institut. We acknowledge financial support from the Israeli Science Foundation grant 249/10 for JB and DE, and grant 666/13 for GD, RO, GB, and AK. The research leading to these results has received funding from the European Community's
Seventh Framework Programme (FP7/2007-2013) CALIPSO under grant agreement no. 312284. We also thank Daniel Podolsky for helpful discussions.


\section*{Appendix: XAS analysis}

\begin{figure}[htbp]
\centering
\epsfig{file=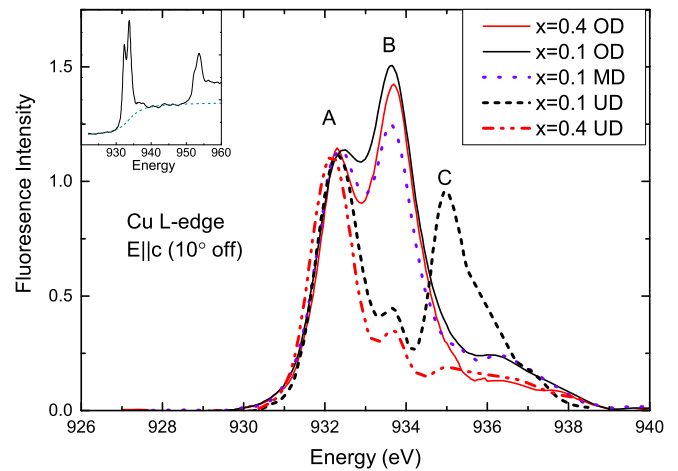,height=2.9in,keepaspectratio}
\caption{(Color Online) The x-ray absorption spectra at the copper $L_{III}$%
-edge, after background subtraction, for the four CLBLCO samples; $x$=0.4
and $x$=0.1 optimally (OD) doped and underdoped (UD) samples, and an
additional $x$=0.1 sample at medium doping (MD). The electric field was
aligned 10$^{\circ}$ from the \textit{c}-axis. The three main peaks are
labeled $A$ and $B$ and $C$. The inset shows an example fitting of the
background.}
\label{fig:xas}
\end{figure}

\begin{figure}[tbp]
	\centering \epsfig{%
		file=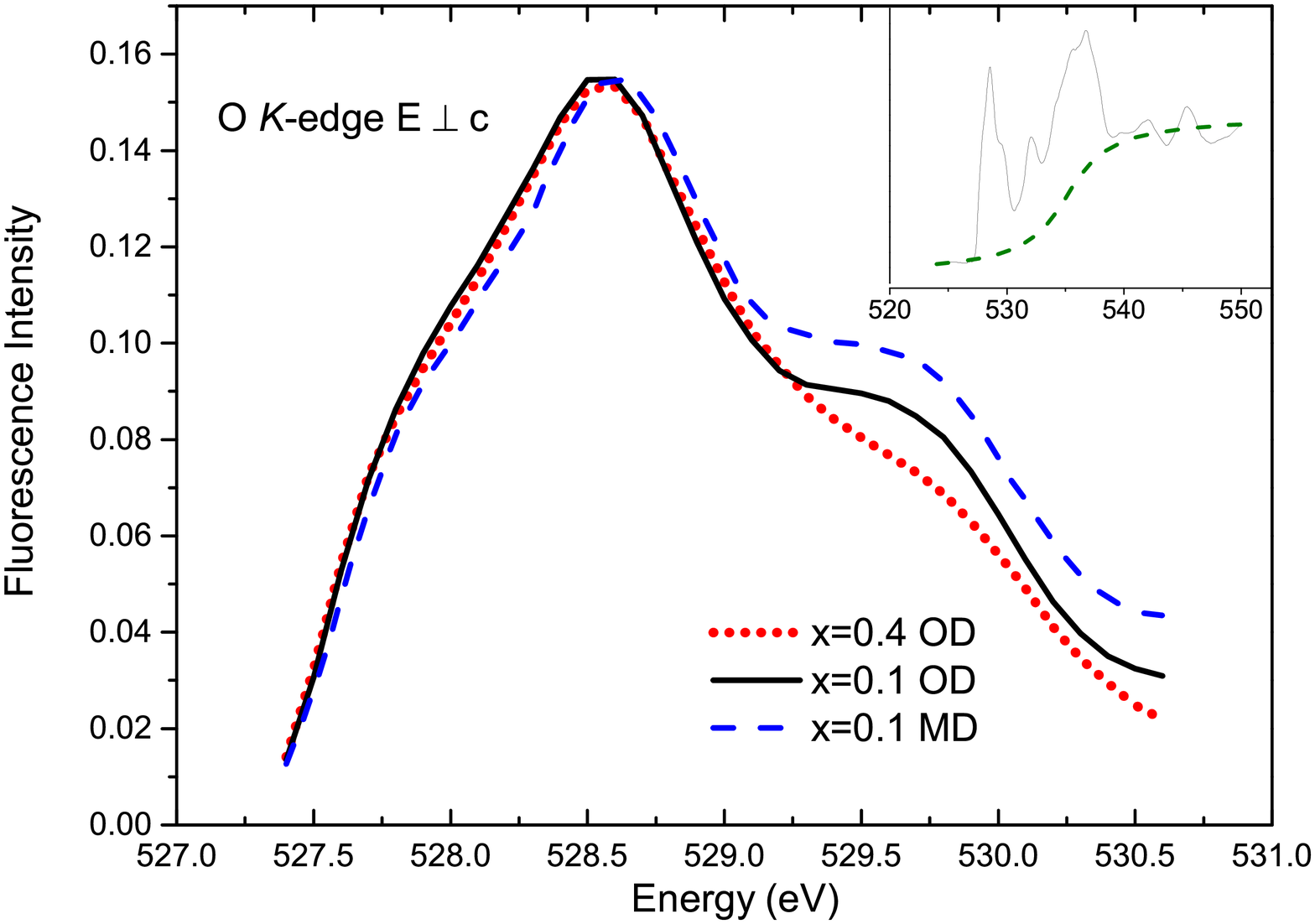,height=2.6in,keepaspectratio} \caption{%
		(Color Online) XAS at the oxygen $K$-edge after background subtraction, with
		electric field parallel to the $ab$ plane. The data were normalized to have
		the same maximum intensity.} 
	\label{fig:xas2} 
\end{figure}

In addition to determining the resonance energy needed for RIXS, XAS also
provides valuable information about the number of holes present in our
samples. We measured the XAS of the single crystal $x$=0.4 OD and UD, and $x$%
=0.1 OD, and UD samples.  In addition, an $x$=0.1 sample of intermediate doping (MD), estimated to be just before the onset of superconductivity, was measured.  We used much of the same approach for analysis
as was used by Agrestini \emph{et al.}\cite{Agrestini14} for treatment of
CLBLCO powder. The clearest and most systematic spectra were at the Cu $%
L_{III}$-edge when the electric field is polarized along the $\mathit{c}$%
-axis (with a 10$^{\circ}$ misalignment from axis), and at the O $K$-edge
absorption when the electric field is polarized parallel to the $ab$-plane.
These are shown in Fig.~\ref{fig:xas} and Fig.~\ref{fig:xas2} respectively,
after subtracting a background in the form of an inverse tangent function as
shown in the insets. Referring to Fig.~\ref{fig:xas}, the data were
normalized so as to have the same maxima of peak $A$ for all samples, which comes from the
Cu $3d^{9}$ $\rightarrow $ Cu $2\bar{p}3d^{10}$ transition \cite{Nucker95,Agrestini14}. 

The low energy edges of the $A$ peaks of all samples
match perfectly, with the exception of $x$=0.4 UD whose $A$ peak is shifted
to slightly lower energy. The second peak $B$ is at the same energy for all
samples. It corresponds to the same absorption process as $A$, but in the
presence of a ligand hole, namely Cu $3d^{9}\bar{L}$ $\rightarrow $ Cu $2%
\bar{p}3d^{10}\bar{L}$\cite{Nucker95,Agrestini14}. It is clear that peak $B$ becomes less and less
intense as the doping decreases, but is roughly the same between $x$=0.1 and
$x$=0.4 for identical nominal dopings. A third peak $C$ appears for the UD
samples $\sim$3 eV from peak $A$. It is quite strong for $x$=0.l but is only
a small bump for $x$=0.4. Such a peak is associated with charge transfer
excitations to the upper Hubbard band\cite{Fink94,Merz98}. A satellite peak around that energy has been related to the chain layer in the
123-compounds\cite{Salluzzo08}.

\begin{figure}[htbp]
	\centering
	\epsfig{file=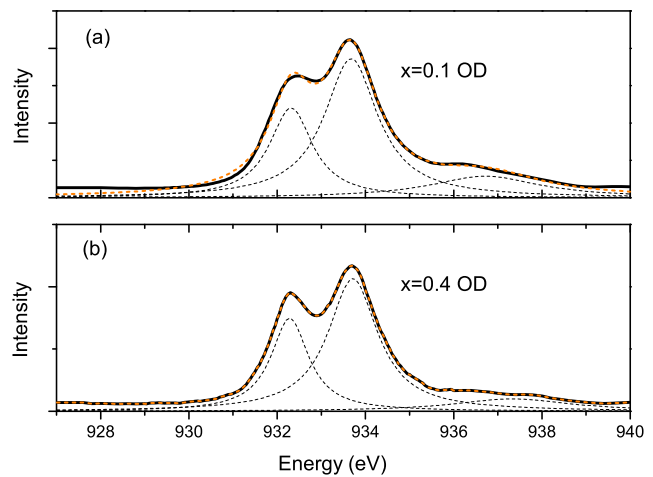,height=2.8in,keepaspectratio}
	\caption{(Color Online) Fitting of the background-subtracted XAS spectra to
		three Lorentzians for the optimally doped samples for (a) $x$=0.1 and (b) $x$%
		=0.4.}
	\label{fig:xasfitting}
\end{figure}

The number of holes can be determined from the relative $B$ peak intensity
\cite{Agrestini14,Kuiper88}. The spectra were fitted to three Lorentzians,
as shown in Fig.~\ref{fig:xasfitting} for the OD samples. The ratio of the
areas of the components, $B/(A+B)$, for OD $x=0.4$ and $x=0.1$ samples were $%
0.652\pm 0.01$ and $0.657\pm 0.01$ respectively, indicating identical hole
doping for both samples. Additionally, we can estimate $y$ and the total number of holes in a unit cell including chains and planes, $h$. Roughly
20\% of $h$ is expected to be in each plane \cite{Nucker95}. From the measured $T_{c}$ of the OD samples, combined with the
phase diagram for powders (see Fig.~\ref{fig:phase_diagram}) \cite{Ofer08}, we obtain $y$=7.06 and 7.11 for the
$x$=0.4 and $x$=0.1 samples respectively. This is near the top, but slightly
to the left of the peak of the superconducting domes. We then estimate the
amount of holes using the relation $h=y-6.25$ \cite{Agrestini14}. Using that as a
reference, and assuming the $B/(A+B)$ area ratios are proportional to $h$,
we can estimate $h$ and $y$ of the UD and MD samples. A summary of the intensity
ratios, estimated $h$, and estimated $y$ for the various samples is
tabulated in table \ref{table:xas}. We note that $y_{UD}\simeq$ 6.32-6.34,
placing it well into the antiferromagnetic long-range ordered phase (Fig.~\ref{fig:phase_diagram}). Likewise, $y_{MD}\simeq
6.94$, which is consistent with the iodometric titration result of 6.92 for
this sample.

\begin{table}
	
	\begin{center}
		
		\caption {Table of parameters determined from the Cu $L$-edge absorption spectra.  The columns are sample, relative area of the $B$ peak, estimated total number of holes $h$, and estimated oxygen content $y$.  As described in the text, for the first two rows $y$ was estimated based on $T_c$, and then $h$ calculated.  In subsequent rows $h$ was calculated first, followed by $y$.}
		
		\begin{ruledtabular}
			
			\begin{tabular}{cccc}
				
				Sample  & $B/(A+B)$ & $h$  & $y$ \\
				
				\hline
				
				$x$=0.4 OD & $0.652\pm 0.01$ & 0.86 & 7.11 \\
				
				$x$=0.1 OD  & $0.657\pm 0.01$  & 0.81 & 7.06 \\
				
				$x$=0.1 MD  & $0.561\pm 0.02$ & 0.69  & 6.94\\
				
				$x$=0.1 UD  & $0.055\pm 0.02$ & 0.07 & 6.32 \\
				
				$x$=0.4 UD  & $0.067\pm 0.01$ & 0.09 & 6.34 \\
				
			\end{tabular}
			
		\end{ruledtabular}
		\label{table:xas}
		
	\end{center}
\end{table}

To further
compare the relative hole densities, the normalized oxygen $K$-edge spectra
is plotted in Fig.~\ref{fig:xas2}. It was measured for the $x$=0.1 OD, $x$%
=0.4 OD, and $x$=0.1 MD samples. 
The effect
of the holes may be seen by inspection of the positions of the low-energy
peak of the O $K$-edge spectra. Shifts in this oxygen ``pre-edge'' energy
track the shift in Fermi level with hole doping \cite{Kuiper88,Nucker95}.
This shift is a direct consequence of the filling (or emptying) of the bands.
From Fig.~\ref{fig:xas2}, the low-energy oxygen $K$-edges overlap almost
exactly for the $x=0.4$ and $x=0.1$ OD samples. In contrast, the edge of the
$x=0.1$ MD spectrum shifts by about 0.07 eV. Based on the result shown for
YBCO in Ref.~52, the shift would correspond to a change in doping of $\delta
y\simeq 0.20$. This is of the same order of magnitude as $\delta y\simeq 0.12
$ between the OD and MD samples in table~\ref{table:xas}. The almost
overlapping edges for the $x$=0.1 and $x$=0.4 OD samples is therefore a
second confirmation of identical number of holes, and furthermore indicates
that the amount holes in the plane layer are the same. 

\vspace{20mm}

\end{document}